\documentclass{Odyssey2026}

\interspeechcameraready 

\usepackage{float}
\usepackage{booktabs}
\usepackage{multirow}
\usepackage{amsmath}
\usepackage{makecell}
\usepackage{adjustbox}

\usepackage{xcolor}
\definecolor{mydarkgray}{RGB}{90,90,90}
\newcommand{\perf}[2]{#1\,{\tiny\textcolor{mydarkgray}{#2}}}

\title{On Low-Bit Quantization Errors in Speaker Verification: Diagnostic and Mitigation}

\author[affiliation={1,2}]{Hugo}{Leguillier}
\author[affiliation={1}]{Driss}{Matrouf}
\author[affiliation={2}]{Guillaume}{Lechien}
\author[affiliation={1}]{Mickael}{Rouvier}

\affiliation{Avignon University}{LIA, UPR 4128}{France}
\email{name.surname@univ-avignon.fr}
\affiliation{}{Aday}{France}
\email{firstname.lastname@univ-avignon.fr, glechien@aday.fr}
\keywords{speaker verification, low-bit quantization, quantization-aware training, score drift, multi-precision cascade}

\usepackage{comment}
\usepackage{soul}
\begin{document}

\maketitle

\begin{abstract}
    
    
Although low-bit quantization provides practical means to deploy speaker verification on resource-constrained devices, its effects on speaker verification performance remain poorly understood. In this paper, we study uniform K-means quantization-aware training of ResNet-36 and ResNet-200 through joint layer-wise and score-level analyses. Our layer-wise analysis highlights fragile components and shows that score degradation is not fully explained by weight distortion alone. We identify a clear knee point at 2 bits, with larger score drift and harmful decision flips concentrated near the FP32 threshold. Our score-level analysis reveals where and how score errors emerge under extreme quantization. Building on these findings, we propose a calibrated multi-precision cascade that resolves most trials at 2 bits and escalates only ambiguous cases, achieving performance close to FP32 while preserving the efficiency benefits of low-bit inference with substantially lower compute and memory costs.
\end{abstract}

%%%%%%%%%%%%%%%%%%%%%%%%%%%%%%%%%%%%%%%%%%%%%%%%%%%%%%%%%%%%%%%%%%%%%%%%
%%%%%%%%%%%%%%%%%%%%%%%%%%% Intro  %%%%%%%%%%%%%%%%%%%%%%%%%%%%%%%%%%%%
%%%%%%%%%%%%%%%%%%%%%%%%%%%%%%%%%%%%%%%%%%%%%%%%%%%%%%%%%%%%%%%%%%%%%%%%
\section{Introduction}

Speaker verification (SV) aims to verify a speaker's claimed identity from their voice. Over the past decade, SV has largely shifted toward deep embedding systems, in which a neural encoder maps variable-length speech utterances to fixed-dimensional speaker representations, referred to as \emph{speaker embeddings}, which are then compared by a scoring backend. Architectures such as ResNet~\cite{zeinali2019but}, ECAPA-TDNN~\cite{desplanques20_interspeech}, ECAPA2~\cite{thienpondt2023ecapa2} and RedimNet~\cite{yakovlev24_interspeech} have significantly improved accuracy and robustness, but at the cost of increased model size and computational complexity.

High-performing SV backbones often require substantial memory and compute resources, which limits their deployment on embedded or resource-constrained devices. This has motivated growing interest in lightweight SV, including compact architectures, asymmetric enroll--verify pipelines~\cite{lin2022lightweightapplicationsasymmetricenrollverify} and model compression techniques such as knowledge distillation and quantization~\cite{wang2019kd_smallfootprint,liu2022self_kd,li23t_interspeech}. Among these approaches, quantization is particularly attractive because it directly reduces the memory footprint and arithmetic cost at deployment without requiring a redesign of the embedding extractor. Recent work has shown that SV models can tolerate moderate quantization, typically 8-bit or 4-bit~\cite{li23t_interspeech,wang23u_interspeech}, and Liu \textit{et al.}~\cite{Liu_2024_KMQAT} further introduced an adaptive K-Means Quantization-Aware Training framework (KMQAT) that constructs layer-specific centroids and extends naturally to mixed-precision quantization and multi-stage fine-tuning (MSFT)~\cite{Liu_2024_KMQAT}. 

However, prior work mostly evaluates quantization through aggregate metrics such as EER, model size or computational cost. While necessary, these metrics do not explain how low-bit degradation develops inside SV systems. In particular, quantization noise may affect internal components unevenly and may induce structured decision errors rather than uniform performance shifts. Moreover, most prior studies evaluate quantized SV systems primarily on in-domain benchmarks and report only limited evidence on domain generalization. While cross-language evaluation on CN-Celeb has been considered in prior model compression work \cite{li23t_interspeech}, comparable analyses across multiple out-of-domain corpora remain rare in speaker verification. In this work, we therefore systematically evaluate all precision regimes not only in-domain, but also on out-of-domain benchmarks, including CN-Celeb and CommonBench, in order to assess the robustness of low-bit quantization under domain shift.

In this paper, we propose a structured analysis of low-bit quantization for speaker verification, with the goal of understanding not only how much performance is lost, but also where this degradation arises and how it affects decision making. Using KMQAT as a controlled experimental framework, we study ResNet-36 and ResNet-200 under uniform 4-, 3-, and 2-bit quantization on both in-domain and out-of-domain benchmarks. We first analyze degradation at the system and block levels to identify the components that are most sensitive to aggressive quantization. We then examine the effect of low-bit quantization in the score space, showing how score drift evolves with precision and how harmful decision flips concentrate near the FP32 operating threshold. Based on these observations, we finally introduce a calibrated multi-precision cascade that resolves most trials at very low precision and escalates only ambiguous cases to higher precision.

The main contributions of this paper are as follows:
\begin{itemize}
    \item We provide a joint layer-wise and score-level analysis of low-bit KMQAT for speaker verification on ResNet-36 and ResNet-200 across the 4-, 3-, and 2-bit regimes.
    \item We perform an evaluation on in-domain VoxCeleb benchmarks but also on out-of-domain corpora, with CN-Celeb and CommonBench, to assess the robustness of low-bit quantization under domain shift.
    \item We identify 2-bit quantization as the main degradation regime and show that quantization sensitivity is highly non-uniform across components and evaluation conditions.
    \item We show that harmful decision errors concentrate near the FP32 operating threshold, while scores remain strongly correlated between precisions.
    \item We leverage these findings to design a calibrated multi-precision cascade that retains most of the efficiency benefits of 2-bit inference while remaining close to FP32 performance.
\end{itemize}

The paper is organized as follows: Section~\ref{sec:KMQAT} introduces the quantization-aware training framework. Section~\ref{sec:Experimental_Protocol} introduces the ResNet architectures and describes the training protocol. Section~\ref{sec:Layer_analysis} studies low-bit degradation from a layer-wise perspective, while Section~\ref{sec:score_analysis} analyzes its impact at the score and decision levels. Section~\ref{sec:Cascading} leverages these findings to design a calibrated multi-precision cascade. Section~\ref{sec:Conclusion} concludes the paper.

%%%%%%%%%%%%%%%%%%%%%%%%%%%%%%%%%%%%%%%%%%%%%%%%%%%%%%%%%%%%%%%%%%%%%%%%
%%%%%%%%%%%%%%%%%%%%%%%%%%% Section II %%%%%%%%%%%%%%%%%%%%%%%%%%%%%%%%
%%%%%%%%%%%%%%%%%%%%%%%%%%%%%%%%%%%%%%%%%%%%%%%%%%%%%%%%%%%%%%%%%%%%%%%%

\section{K-Means Quantization-Aware Training}
\label{sec:KMQAT}

We build on the K-Means Quantization-Aware Training (KMQAT) framework introduced by Liu \textit{et al.} \cite{Liu_2024_KMQAT}, and use a uniform layer-wise version throughout this study. The goal is to replace the FP32 weights of each layer by a small set of learned quantization levels while preserving the verification performance of the speaker embedding model.

Let $W_l$ denote the FP32 weight tensor of layer $l$. For $b$-bit quantization, KMQAT associates this layer with a codebook:
\[
\mathcal{C}_l = \{c_{l,1}, \dots, c_{l,K}\}, \qquad K = 2^b,
\]

Quantized weights are represented by centroids obtained by a K-means-like clustering procedure, together with a learned scaling factor. In the layer-wise setting considered here, each layer uses a single shared codebook and a single learned scale $\alpha_l$.

The codebook is initialized from the pretrained FP32 weights. Following Liu \textit{et al.}~\cite{Liu_2024_KMQAT}, we first discard extreme outliers by retaining only the central portion of the weight distribution. The clipped weights are then partitioned into $K$ intervals, and one centroid is initialized per interval as the empirical mean of the weights falling in that interval. The resulting centroids are finally rescaled to a normalized range before quantization-aware training. This initialization provides layer-specific quantization levels that adapt to the statistics of each weight tensor.

Given $\mathcal{C}_l$ and $\alpha_l$, each scalar weight $w_l \in W_l$ is first mapped to the normalized space and assigned to its nearest centroid:
\begin{equation}
q_{l}
=
\arg\min_{c \in \mathcal{C}_l}
\left|
\frac{w_{l}}{\alpha_l} - c
\right|.
\label{eq:kmqat_assign}
\end{equation}
The quantized scalar weight is then reconstructed as
\begin{equation}
\hat{w}_{l} = \alpha_l q_{l},
\label{eq:kmqat_scalar}
\end{equation}
which yields the quantized tensor
\begin{equation}
\widehat{W}_l = \alpha_l Q_l,
\label{eq:kmqat_tensor}
\end{equation}
where $Q_l$ denotes the tensor obtained by stacking the assigned centroids $q_{l}$ at the same positions as the original weights in $W_l$.

Training is performed with quantization-aware training. The discrete assignment is handled with a straight-through estimator (STE) \cite{bengio2013STE}, which allows gradients to flow to the underlying FP32 weights and to the learned scaling factors.
In addition, we periodically reassign centroid indices during training so that the discrete proxy remains consistent with the evolving FP32 weights. This reassignment plays the role of an online refinement of the quantized representation during QAT. In practice, we apply this procedure to uniform 4-, 3-, and 2-bit quantization of the ResNet backbones studied in this paper.

We use this controlled KMQAT setting as an experimental method for analyzing low-bit degradation in speaker verification. The layer-wise and score-level analyses in the following sections are therefore performed under a common quantization framework, which enables direct comparisons across architectures and different bit-widths.

%%%%%%%%%%%%%%%%%%%%%%%%%%%%%%%%%%%%%%%%%%%%%%%%%%%%%%%%%%%%%%%%%%%%%%%%
%%%%%%%%%%%%%%%%%%%%%%%%%%% Section III %%%%%%%%%%%%%%%%%%%%%%%%%%%%%%%%
%%%%%%%%%%%%%%%%%%%%%%%%%%%%%%%%%%%%%%%%%%%%%%%%%%%%%%%%%%%%%%%%%%%%%%%%
\begin{table*}[t]
\centering
\caption{Speaker verification performance of FP32 and uniformly quantized ResNet-36 and ResNet-200 models on in-domain and out-of-domain benchmarks. Quantized models use 4-, 3-, or 2-bit precision throughout the network. Relative differences are computed with respect to the FP32 baseline of the same architecture. Bold values indicate the best quantized result for each architecture and metric, excluding the FP32 reference.}
\label{tab:quant_results}
\renewcommand{\arraystretch}{1.15}
\setlength{\tabcolsep}{4pt}

\begin{adjustbox}{width=\textwidth}
\begin{tabular}{lcll|llllllll|llllllll}
\toprule
\multirow{3}{*}{\textbf{System}} 
& \multirow{3}{*}{\textbf{\makecell{Bit Width\\(bit)}}}
& \multicolumn{2}{c}{\textbf{Overall}}
& \multicolumn{8}{c}{\textbf{In-Domain}}
& \multicolumn{6}{c}{\textbf{Out-of-Domain}} \\
\cmidrule(lr){3-4} \cmidrule(lr){5-12} \cmidrule(lr){13-18}
& &
\multicolumn{2}{c}{\textbf{Avg.}}
& \multicolumn{2}{c}{\textbf{Avg.}}
& \multicolumn{2}{c}{\textbf{Vox1-O}}
& \multicolumn{2}{c}{\textbf{Vox1-E}}
& \multicolumn{2}{c}{\textbf{Vox1-H}}
& \multicolumn{2}{c}{\textbf{Avg.}}
& \multicolumn{2}{c}{\textbf{CommonBench}}
& \multicolumn{2}{c}{\textbf{CN-Celeb}} \\
\cmidrule(lr){3-4} \cmidrule(lr){5-6} \cmidrule(lr){7-8} \cmidrule(lr){9-10} \cmidrule(lr){11-12} \cmidrule(lr){13-14} \cmidrule(lr){15-16} \cmidrule(lr){17-18}
& &
\multicolumn{1}{c}{\textbf{EER}} & \multicolumn{1}{c}{\textbf{minDCF}}
& \multicolumn{1}{c}{\textbf{EER}} & \multicolumn{1}{c}{\textbf{minDCF}}
& \multicolumn{1}{c}{\textbf{EER}} & \multicolumn{1}{c}{\textbf{minDCF}}
& \multicolumn{1}{c}{\textbf{EER}} & \multicolumn{1}{c}{\textbf{minDCF}}
& \multicolumn{1}{c}{\textbf{EER}} & \multicolumn{1}{c}{\textbf{minDCF}}
& \multicolumn{1}{c}{\textbf{EER}} & \multicolumn{1}{c}{\textbf{minDCF}}
& \multicolumn{1}{c}{\textbf{EER}} & \multicolumn{1}{c}{\textbf{minDCF}}
& \multicolumn{1}{c}{\textbf{EER}} & \multicolumn{1}{c}{\textbf{minDCF}} \\
\midrule

\multirow{4}{*}{ResNet-36}
& 32 
& 3.909
& 0.254
& 1.234
& 0.118
& 0.973
& 0.085
& 0.994
& 0.102
& 1.736
& 0.167
& 7.923
& 0.458
& 3.922
& 0.360
& 11.923
& 0.555 \\

& 4 
& \perf{\textbf{3.949}}{+1.0\%} 
& \perf{\textbf{0.263}}{+3.5\%}
& \perf{\textbf{1.278}}{+3.6\%} 
& \perf{0.124}{+5.1\%}
& \perf{\textbf{0.994}}{+2.2\%} 
& \perf{0.0856}{+0.7\%}
& \perf{\textbf{1.042}}{+4.8\%} 
& \perf{\textbf{0.110}}{+7.8\%}
& \perf{\textbf{1.799}}{+3.6\%} 
& \perf{\textbf{0.175}}{+4.8\%}
& \perf{\textbf{7.955}}{+0.4\%} 
& \perf{\textbf{0.471}}{+2.8\%}
& \perf{\textbf{4.077}}{+4.0\%} 
& \perf{\textbf{0.376}}{+4.4\%}
& \perf{\textbf{11.833}}{-0.8\%} 
& \perf{\textbf{0.566}}{+2.0\%} \\

& 3 
& \perf{3.965}{+1.4\%} 
& \perf{\textbf{0.263}}{+3.5\%}
& \perf{1.297}{+5.1\%} 
& \perf{\textbf{0.123}}{+4.2\%}
& \perf{0.999}{+2.7\%} 
& \perf{\textbf{0.0794}}{-6.6\%}
& \perf{1.053}{+5.9\%} 
& \perf{0.113}{+10.8\%}
& \perf{1.840}{+6.0\%} 
& \perf{0.176}{+5.4\%}
& \perf{7.966}{+0.5\%} 
& \perf{0.473}{+3.3\%}
& \perf{4.078}{+4.0\%} 
& \perf{0.379}{+5.3\%}
& \perf{11.855}{-0.6\%} 
& \perf{0.568}{+2.3\%} \\

& 2 
& \perf{4.292}{+9.8\%} 
& \perf{0.287}{+13.0\%}
& \perf{1.462}{+18.5\%} 
& \perf{0.146}{+23.7\%}
& \perf{1.068}{+9.8\%} 
& \perf{0.1102}{+29.6\%}
& \perf{1.229}{+23.6\%} 
& \perf{0.135}{+32.4\%}
& \perf{2.089}{+20.3\%} 
& \perf{0.193}{+15.6\%}
& \perf{8.538}{+7.8\%} 
& \perf{0.499}{+9.0\%}
& \perf{4.568}{+16.5\%} 
& \perf{0.407}{+13.1\%}
& \perf{12.509}{+4.9\%} 
& \perf{0.592}{+6.7\%} \\
\midrule

\multirow{4}{*}{ResNet-200}
& 32 
& 3.443
& 0.224
& 0.984
& 0.091
& 0.715
& 0.053
& 0.820
& 0.083
& 1.418
& 0.136
& 7.131
& 0.424
& 3.456
& 0.329
& 10.807
& 0.519 \\

& 4 
& \perf{\textbf{3.637}}{+5.6\%} 
& \perf{0.241}{+7.6\%}
& \perf{\textbf{1.076}}{+9.3\%} 
& \perf{\textbf{0.102}}{+12.1\%}
& \perf{0.834}{+16.6\%} 
& \perf{0.0655}{+23.6\%}
& \perf{\textbf{0.880}}{+7.3\%} 
& \perf{\textbf{0.094}}{+13.3\%}
& \perf{\textbf{1.514}}{+6.8\%} 
& \perf{\textbf{0.147}}{+8.1\%}
& \perf{\textbf{7.479}}{+4.9\%} 
& \perf{0.450}{+6.1\%}
& \perf{\textbf{3.603}}{+4.3\%} 
& \perf{\textbf{0.345}}{+4.9\%}
& \perf{11.355}{+5.1\%} 
& \perf{0.554}{+6.7\%} \\

& 3 
& \perf{3.643}{+5.8\%} 
& \perf{\textbf{0.240}}{+7.1\%}
& \perf{1.081}{+9.9\%} 
& \perf{0.102}{+12.1\%}
& \perf{\textbf{0.789}}{+10.3\%} 
& \perf{\textbf{0.0620}}{+17.0\%}
& \perf{0.900}{+9.8\%} 
& \perf{0.096}{+15.7\%}
& \perf{1.554}{+9.6\%} 
& \perf{\textbf{0.147}}{+8.1\%}
& \perf{7.487}{+5.0\%} 
& \perf{\textbf{0.447}}{+5.4\%}
& \perf{3.626}{+4.9\%} 
& \perf{\textbf{0.345}}{+4.9\%}
& \perf{11.349}{+5.0\%} 
& \perf{\textbf{0.548}}{+5.6\%} \\

& 2 
& \perf{3.685}{+7.0\%} 
& \perf{0.253}{+12.9\%}
& \perf{1.127}{+14.5\%} 
& \perf{0.112}{+23.1\%}
& \perf{0.813}{+13.7\%} 
& \perf{0.0824}{+55.5\%}
& \perf{0.946}{+15.4\%} 
& \perf{0.098}{+18.1\%}
& \perf{1.623}{+14.5\%} 
& \perf{0.157}{+15.4\%}
& \perf{7.523}{+5.5\%} 
& \perf{0.464}{+9.4\%}
& \perf{3.732}{+8.0\%} 
& \perf{0.357}{+8.5\%}
& \perf{\textbf{11.314}}{+4.7\%} 
& \perf{0.571}{+10.0\%} \\
\bottomrule
\end{tabular}
\end{adjustbox}
\end{table*}

\section{Experimental Protocol}
\label{sec:Experimental_Protocol}

We first describe the training setup used for both FP32 and quantized models (Section~\ref{sec:training_setup}), then introduce the ResNet architectures considered in the study (Section~\ref{sec:resnet_architectures}), and finally present the evaluation protocol used for in-domain and out-of-domain assessment (Section~\ref{sec:evaluation_protocol}).

\subsection{Training setup}
\label{sec:training_setup}

All experiments are conducted using the Kiwano toolkit~\cite{rouvier2026}\footnote{\url{https://github.com/kiwano-toolkit/kiwano}}. The framework implements the complete training, quantization, embedding extraction, and scoring pipeline used in this work. This also improves reproducibility with respect to prior KMQAT/MSFT publications, which introduced the method and training strategy but, to the best of our knowledge, did not provide a public code release referenced in the paper versions. We train ResNet-based speaker embedding extractors on VoxCeleb2~\cite{chung2018voxceleb2}. Input features are 80-dimensional log Mel filterbanks, and the embedding dimension is set to 256. During training, each mini-batch contains 512 randomly cropped speech segments of 3.5\,s. Standard data augmentation is applied following~\cite{snyder2018x}, using MUSAN~\cite{snyder2015musanmusicspeechnoise} and simulated room impulse responses~\cite{rirs}.

Models are trained for 42 epochs with Additive Angular Margin (AAM) loss, using SGD with momentum $0.9$ and weight decay $2 \times 10^{-4}$. Following~\cite{rouvier2021studyingsqueezeandexcitationusedcnn}, Squeeze-and-Excitation modules are enabled in the first two stages. The resulting FP32 models are then used as initialization for quantization-aware training.

For quantized experiments, KMQAT is initialized from the corresponding pretrained FP32 checkpoints and the models are fine-tuned for 40 epochs. We fine-tune separate models for 4, 3, and 2 bits. The training objective remains the same as for the FP32 model, namely the AAM loss; only the weights are quantized during the forward pass through the QAT procedure. All quantized runs use a retention ratio of $r=0.9$ to discard the outer 10\% weight outliers before codebook initialization, together with symmetric centroid rescaling, and an equal-mass partition of the retained weight distribution. Data augmentation is disabled during this phase.

In addition to these uniform models, we also train mixed-precision models using the multi-stage fine-tuning (MSFT) strategy introduced by Liu \textit{et al.} \cite{Liu_2024_KMQAT}, where layers are quantized progressively according to a predefined bit allocation in order to stabilize aggressive low-bit adaptation.

\subsection{ResNet architectures}
\label{sec:resnet_architectures}

We consider two ResNet backbones, denoted \textit{ResNet-36} and \textit{ResNet-200}. Both models share the same stage widths, namely $\{128, 128, 256, 256\}$ feature maps, and the same residual block design. They differ only in the number of residual blocks per stage. Following the convention used throughout this paper, the configuration is given as the number of blocks in stages 1 to 4. ResNet-36 uses the configuration $[3, 4, 6, 3]$, whereas ResNet-200 uses $[3, 24, 36, 3]$. This controlled setup allows us to compare the behavior of low-bit quantization across two architectures that differ primarily in depth.

\subsection{Evaluation protocol}
\label{sec:evaluation_protocol}

We evaluate the systems on VoxCeleb1-O/E/H (cleaned) test sets~\cite{Nagrani17}, which serve as in-domain benchmarks. To assess robustness beyond the training domain, we additionally report results on the out-of-domain benchmarks CommonBench~\cite{hintz24_spsc} and CN-Celeb~\cite{fan2019cnceleb}. Trial scores are computed using cosine similarity between speaker embeddings.

We report Equal Error Rate (EER) and minimum Detection Cost Function (minDCF), with $P_{\mathrm{tar}} = 10^{-2}$ and $C_{\mathrm{miss}} = C_{\mathrm{fa}} = 1$.

%%%%%%%%%%%%%%%%%%%%%%%%%%%%%%%%%%%%%%%%%%%%%%%%%%%%%%%%%%%%%%%%%%%%%%%%
%%%%%%%%%%%%%%%%%%%%%%%%%%% Section IV %%%%%%%%%%%%%%%%%%%%%%%%%%%%%%%%
%%%%%%%%%%%%%%%%%%%%%%%%%%%%%%%%%%%%%%%%%%%%%%%%%%%%%%%%%%%%%%%%%%%%%%%%

\section{Layer-wise Analysis}
\label{sec:Layer_analysis}

In this section, we first study the impact of uniform quantization across bit-widths to determine the regime in which performance loss becomes substantial (Section~\ref{sec:global_sv_degradation}). We then focus on this regime and perform a stage-wise sensitivity analysis to identify the components that are most critical under aggressive quantization (Section~\ref{sec:internal_diagnostics_exclusion}).

\subsection{System-level degradation under uniform quantization}
\label{sec:global_sv_degradation}

We first evaluate uniform low-bit quantization at the system level to determine whether a finer-grained internal analysis is justified. If speaker verification performance remained essentially stable across bit-widths, then investigating layer-wise sensitivity would be of limited practical interest. Table~\ref{tab:quant_results} compares the FP32 ResNet-36 and ResNet-200 baselines with their uniformly quantized 4-, 3-, and 2-bit counterparts on both in-domain benchmarks (VoxCeleb1-O/E/H) and out-of-domain benchmarks (CommonBench and CN-Celeb). In addition to the EER and minDCF, the table reports the relative variation with respect to the corresponding 32-bit model, providing a clearer view of the impact of quantization across evaluation conditions.

A consistent pattern is observed across both architectures. Performance degradation remains limited at 4 bits, becomes slightly more noticeable at 3 bits, and grows substantially at 2 bits. This trend is particularly clear on the in-domain benchmarks. For example, the average in-domain EER of ResNet-36 increases from 1.234\% in FP32 to 1.278\% at 4 bits, 1.297\% at 3 bits, and 1.462\% at 2 bits, while the corresponding minDCF rises from 0.118 to 0.124, 0.123, and 0.146. ResNet-200 exhibits the same overall pattern, increasing from 0.984\% in FP32 to 1.076\%, 1.081\%, and 1.127\% respectively. Among the in-domain test sets, VoxCeleb1-H is consistently the most affected, indicating that quantization errors become more harmful under the most challenging verification conditions.

The out-of-domain benchmarks confirm the same overall trend, although with more heterogeneous behavior across corpora. In absolute terms, these evaluations are already much more difficult than the in-domain ones, even in FP32. Quantization induces a progressive deterioration overall, especially in the 2-bit regime. On CommonBench, both architectures show a steady increase in EER and minDCF as the bit-width decreases. On CN-Celeb, the effect is more nuanced: for ResNet-36, the EER remains close to the FP32 baseline at 4 and 3 bits, whereas minDCF still increases, suggesting that quantization may affect score calibration even when its effect on discrimination remains limited. However, at 2 bits, the degradation becomes clearly visible on both out-of-domain corpora.

Taken together, these results show that, although ResNet-200 remains stronger than ResNet-36 across both in-domain and out-of-domain evaluations, both models follow the same global pattern: 4-bit and 3-bit quantization remain relatively close to FP32, whereas 2-bit quantization clearly constitutes the main degradation regime. This observation motivates the remainder of the analysis, which focuses on the 2-bit setting in order to identify the internal components that contribute most to the observed performance loss and to examine whether selective retention of higher precision can mitigate it.
\subsection{Stage-wise sensitivity under selective full-precision retention}
\label{sec:internal_diagnostics_exclusion}

\begin{table*}[t]
\centering
\caption{Stage-wise sensitivity analysis under 2-bit quantization. All components of the network are quantized to 2 bits except the stage specified in the \textit{FP32 Stage} column, which remains in FP32. The row marked ``-'' denotes the uniform 2-bit baseline. Bold and underlined values indicate the best and second-best selective-retention results. Performance is reported on in-domain and out-of-domain benchmarks using EER and minDCF.}
\label{tab:block_wise}
\renewcommand{\arraystretch}{1.15}
\setlength{\tabcolsep}{4pt}

\begin{adjustbox}{width=\textwidth}
\begin{tabular}{lcccc|cccccccc|ccccccc}
\toprule
\multirow{3}{*}{\textbf{System}}
& \multirow{3}{*}{\textbf{\makecell{Bit Width\\(bit)}}}
& \multirow{3}{*}{\textbf{FP32 Stage}}
& \multicolumn{2}{c}{\textbf{Overall}}
& \multicolumn{8}{c}{\textbf{In-Domain}}
& \multicolumn{6}{c}{\textbf{Out-of-Domain}} \\
\cmidrule(lr){4-5} \cmidrule(lr){6-13} \cmidrule(lr){14-19}
& & 
& \multicolumn{2}{c}{\textbf{Avg.}}
& \multicolumn{2}{c}{\textbf{Avg.}}
& \multicolumn{2}{c}{\textbf{Vox1-O}}
& \multicolumn{2}{c}{\textbf{Vox1-E}}
& \multicolumn{2}{c}{\textbf{Vox1-H}}
& \multicolumn{2}{c}{\textbf{Avg.}}
& \multicolumn{2}{c}{\textbf{CommonBench}}
& \multicolumn{2}{c}{\textbf{CN-Celeb}} \\
\cmidrule(lr){4-5} \cmidrule(lr){6-7} \cmidrule(lr){8-9} \cmidrule(lr){10-11} \cmidrule(lr){12-13} \cmidrule(lr){14-15} \cmidrule(lr){16-17} \cmidrule(lr){18-19}
& & 
& \textbf{EER} & \textbf{minDCF}
& \textbf{EER} & \textbf{minDCF}
& \textbf{EER} & \textbf{minDCF}
& \textbf{EER} & \textbf{minDCF}
& \textbf{EER} & \textbf{minDCF}
& \textbf{EER} & \textbf{minDCF}
& \textbf{EER} & \textbf{minDCF}
& \textbf{EER} & \textbf{minDCF} \\
\midrule

\multirow{6}{*}{ResNet-36}
& 2  & - 
& 4.292 & 0.287 
& 1.462 & 0.146 
& 1.068 & 0.1102 
& 1.229 & 0.135 
& 2.089 & 0.193 
& 8.538 & 0.499 
& 4.568 & 0.407 
& 12.509 & 0.592 \\

& 2 & Stage 1
& 4.099 & 0.269
& 1.388 & 0.131
& 1.116 & 0.0876
& 1.117 & 0.118
& 1.931 & 0.186
& 8.167 & 0.476
& 4.174 & 0.382
& 12.160 & \underline{0.570} \\

& 2 & Stage 2
& \textbf{3.981} & \textbf{0.266}
& 1.336 & 0.129
& \underline{1.002} & 0.0878
& 1.100 & \underline{0.114}
& 1.906 & 0.185
& \textbf{7.950} & \textbf{0.471}
& \underline{4.094} & 0.379
& \textbf{11.806} & \textbf{0.564} \\

& 2 & Stage 3
& \underline{4.028} & \underline{0.268}
& \textbf{1.323} & 0.130
& 1.028 & 0.0913
& \textbf{1.072} & 0.115
& \textbf{1.869} & \underline{0.184}
& 8.087 & 0.474
& \textbf{4.058} & \textbf{0.373}
& 12.115 & 0.576 \\

& 2 & Stage 4
& 4.038 & \textbf{0.266}
& 1.345 & \underline{0.128}
& 1.031 & \underline{0.0874}
& 1.098 & \textbf{0.113}
& 1.906 & \textbf{0.183}
& \underline{8.078} & \underline{0.472}
& 4.120 & \textbf{0.373}
& \underline{12.036} & 0.572 \\

& 2 & Embedding
& 4.034 & \underline{0.268}
& \underline{1.325} & \textbf{0.126}
& \textbf{0.991} & \textbf{0.0763}
& 1.092 & 0.119
& \underline{1.891} & \textbf{0.183}
& 8.099 & 0.481
& 4.118 & 0.382
& 12.081 & 0.581 \\
\midrule

\multirow{6}{*}{ResNet-200}
& 2 & - 
& 3.685
& 0.253
& 1.127
& 0.112
& 0.813
& 0.0824
& 0.946
& \underline{0.098}
& 1.623
& 0.157
& 7.523
& 0.464
& 3.732
& 0.357
& 11.314
& 0.571 \\

& 2 & Stage 1
& 3.732 & 0.251
& 1.145 & 0.110
& 0.869 & 0.0794
& 0.952 & 0.101
& 1.615 & \underline{0.151}
& 7.613 & 0.462
& 3.686 & 0.356
& 11.540 & 0.568 \\

& 2 & Stage 2
& \underline{3.604} & \underline{0.244}
& 1.089 & \underline{0.108}
& \textbf{0.736} & \underline{0.0719}
& 0.939 & 0.099
& 1.593 & 0.153
& \underline{7.376} & \textbf{0.447}
& \underline{3.657} & \underline{0.347}
& \textbf{11.095} & \textbf{0.547} \\

& 2 & Stage 3
& \textbf{3.589} & \textbf{0.243}
& \textbf{1.080} & \textbf{0.106}
& 0.816 & 0.0732
& \textbf{0.884} & \textbf{0.097}
& \textbf{1.541} & \textbf{0.148}
& \textbf{7.353} & \underline{0.448}
& \textbf{3.593} & \textbf{0.345}
& \underline{11.112} & \underline{0.550} \\

& 2 & Stage 4
& 3.754 & 0.249
& 1.129 & \underline{0.108}
& 0.848 & \textbf{0.0672}
& 0.933 & 0.101
& 1.607 & 0.156
& 7.693 & 0.460
& 3.699 & 0.353
& 11.687 & 0.568 \\

& 2 & Embedding
& 3.704 & 0.250
& \underline{1.086} & 0.109
& \underline{0.763} & 0.0754
& \underline{0.915} & \underline{0.098}
& \underline{1.579} & 0.153
& 7.633 & 0.462
& 3.692 & 0.353
& 11.574 & 0.571 \\
\bottomrule
\end{tabular}
\end{adjustbox}
\end{table*}

To localize the origin of the degradation observed at 2 bits, we consider a selective retention setting in which the whole network is quantized to 2 bits except for one component kept in FP32. In Table~\ref{tab:block_wise}, the \textit{FP32 Stage} column specifies which stage remains in FP32, while all other components are quantized to 2 bits. The row marked ``-'' denotes the reference fully quantized 2-bit baseline, where no component is kept in FP32. Under this protocol, larger performance recovery indicates that the retained component is more sensitive to low-bit quantization and contributes more strongly to the observed degradation.

For ResNet-36, the most sensitive components differ slightly depending on the evaluation condition. On the overall average, the largest recovery is obtained by keeping Stage~2 in FP32, followed by Stage~3. On the in-domain benchmarks, however, Stage~3 becomes the most critical component, with the embedding layer ranking second. On the out-of-domain benchmarks, the strongest recovery is obtained with Stage~2, followed by Stage~4. This indicates that, for the shallower architecture, quantization sensitivity is distributed across several middle-to-late stages and varies with the evaluation domain.

For ResNet-200, the largest recoveries are obtained when retaining the middle stages in FP32, especially Stage~3 followed by Stage~2 on the global average and on the out-of-domain benchmarks. On the in-domain benchmarks, Stage~3 remains the dominant component, while the embedding layer ranks second. Since Stages~2 and~3 contain most of the residual blocks in this architecture, their sensitivity likely reflects both their representational importance and their larger contribution to the total number of quantized parameters. A similar tendency is observed for ResNet-36, where Stages~2 and~3 also provide among the strongest recoveries. Overall, both architectures indicate that the degradation induced by 2-bit quantization is mainly associated with the larger middle stages rather than with the earliest feature-extraction stage.

Two additional observations are worth noting. First, Stage~1 is never among the most sensitive components for either architecture, suggesting that early low-level feature extraction is relatively robust to 2-bit quantization. Second, the fact that the embedding layer is consistently important in-domain for both models highlights the role of the final speaker representation under matched conditions. Taken together, these results show that the impact of 2-bit quantization is highly non-uniform across the network and is mainly driven by a small subset of middle-to-late blocks, which makes them natural candidates for mixed-precision retention.

%%%%%%%%%%%%%%%%%%%%%%%%%%%%%%%%%%%%%%%%%%%%%%%%%%%%%%%%%%%%%%%%%%%%%%%%
%%%%%%%%%%%%%%%%%%%%%%%%%%% Section V %%%%%%%%%%%%%%%%%%%%%%%%%%%%%%%%
%%%%%%%%%%%%%%%%%%%%%%%%%%%%%%%%%%%%%%%%%%%%%%%%%%%%%%%%%%%%%%%%%%%%%%%%

\section{Score-Level Analysis}
\label{sec:score_analysis}

In this section, we study the effect of quantization in the score space. We first characterize the global score drift induced by quantization and analyze how its magnitude evolves as the bit-width decreases (Section~\ref{global_score_drift}). We then examine how these score perturbations affect decision robustness by studying harmful decision flips with respect to the FP32 operating threshold (Section~\ref{sec:harmful}).

\subsection{Global score drift under low-bit quantization}
\label{global_score_drift}

Low-bit quantization can affect speaker verification scores by perturbing the embeddings used to compute cosine similarity. We therefore begin by examining how quantization alters scores with respect to the FP32 reference system. In the remainder of this section, we use the term \emph{drift} to denote the score difference induced by quantization relative to FP32. For a given trial, this quantity measures how much the score changes when embeddings are extracted with a quantized model instead of the FP32 model.

Let $s_{ff}$ denote the score obtained when both enrollment and test embeddings are extracted with the FP32 model, and let $s_{qq}$ denote the score obtained when both sides are extracted with the quantized model. The full low-bit score drift is defined as
\begin{equation}
\Delta s = s_{qq} - s_{ff}.
\label{eq:delta_s}
\end{equation}
To isolate the effect of quantizing only the test side while keeping enrollment in FP32, we also consider
\begin{equation}
\Delta s_{\mathrm{te}} = s_{fq} - s_{ff},
\label{eq:delta_en_te}
\end{equation}
where $s_{fq}$ denotes the score obtained with FP32 enrollment and quantized test embeddings.
Table~\ref{tab:score_drift_interaction} summarizes the main score-level drift statistics. The quantity mean$|\Delta s|$ denotes the mean absolute full drift, while mean$|\Delta s_{\mathrm{te}}|$ denotes the mean absolute test-only drift.

\begin{table}[H]
\centering
\small
\caption{Mean absolute score drift under low-bit quantization for in-domain and out-of-domain evaluations. Results are averaged over VoxCeleb1-O/E/H for the in-domain setting; out-of-domain statistics over CommonBench and CN-Celeb.} \label{tab:score_drift_interaction} \resizebox{\columnwidth}{!}{\begin{tabular}{@{}lccccc@{}}
\toprule
\multirow{2}{*}{\textbf{Model}} & \multirow{2}{*}{\textbf{Bits}} & \multicolumn{2}{c}{\textbf{In-domain}} & \multicolumn{2}{c}{\textbf{Out-of-domain}} \\ \cmidrule(l){3-6} 
 &  & \textbf{mean$|\Delta s|$} & \textbf{mean$|\Delta s_{\mathrm{te}}|$} & \textbf{mean$|\Delta s|$} & \textbf{mean$|\Delta s_{\mathrm{te}}|$} \\ \midrule
\multirow{3}{*}{ResNet-36} & 4 & 0.0147 & 0.0108 & 0.0228 & 0.0174 \\
 & 3 & 0.0175 & 0.0133 & 0.0248 & 0.0190 \\
 & 2 & 0.0301 & 0.0277 & 0.0372 & 0.0327 \\ \midrule
\multirow{3}{*}{ResNet-200} & 4 & 0.0198 & 0.0151 & 0.0291 & 0.0228 \\
 & 3 & 0.0214 & 0.0162 & 0.0308 & 0.0242 \\
 & 2 & 0.0257 & 0.0209 & 0.0353 & 0.0285 \\ \bottomrule
\end{tabular} } \end{table}

A clear trend emerges across both architectures: score drift increases as precision decreases, with a pronounced knee point at 2 bits. For ResNet-36, the mean absolute full drift increases from $0.0147$ at 4 bits to $0.0175$ at 3 bits and $0.0301$ at 2 bits, corresponding to a $2.05\times$ increase between 4 and 2 bits. In the test-only setting, the increase is even sharper, from $0.0108$ to $0.0133$ and $0.0277$, i.e., a $2.58\times$ increase from 4 to 2 bits. ResNet-200 follows the same overall pattern, but more moderately: the mean absolute full drift increases from $0.0198$ at 4 bits to $0.0214$ at 3 bits and $0.0257$ at 2 bits, while the mean absolute test-only drift rises from $0.0151$ to $0.0162$ and $0.0209$. These results confirm at the score level what was already observed at the verification level: the 2-bit regime corresponds to a clear transition toward substantially stronger perturbations.

\begin{figure}[t]
    \centering
    \includegraphics[width=0.80\columnwidth,trim=5pt 5pt 5pt 5pt,
    clip
]{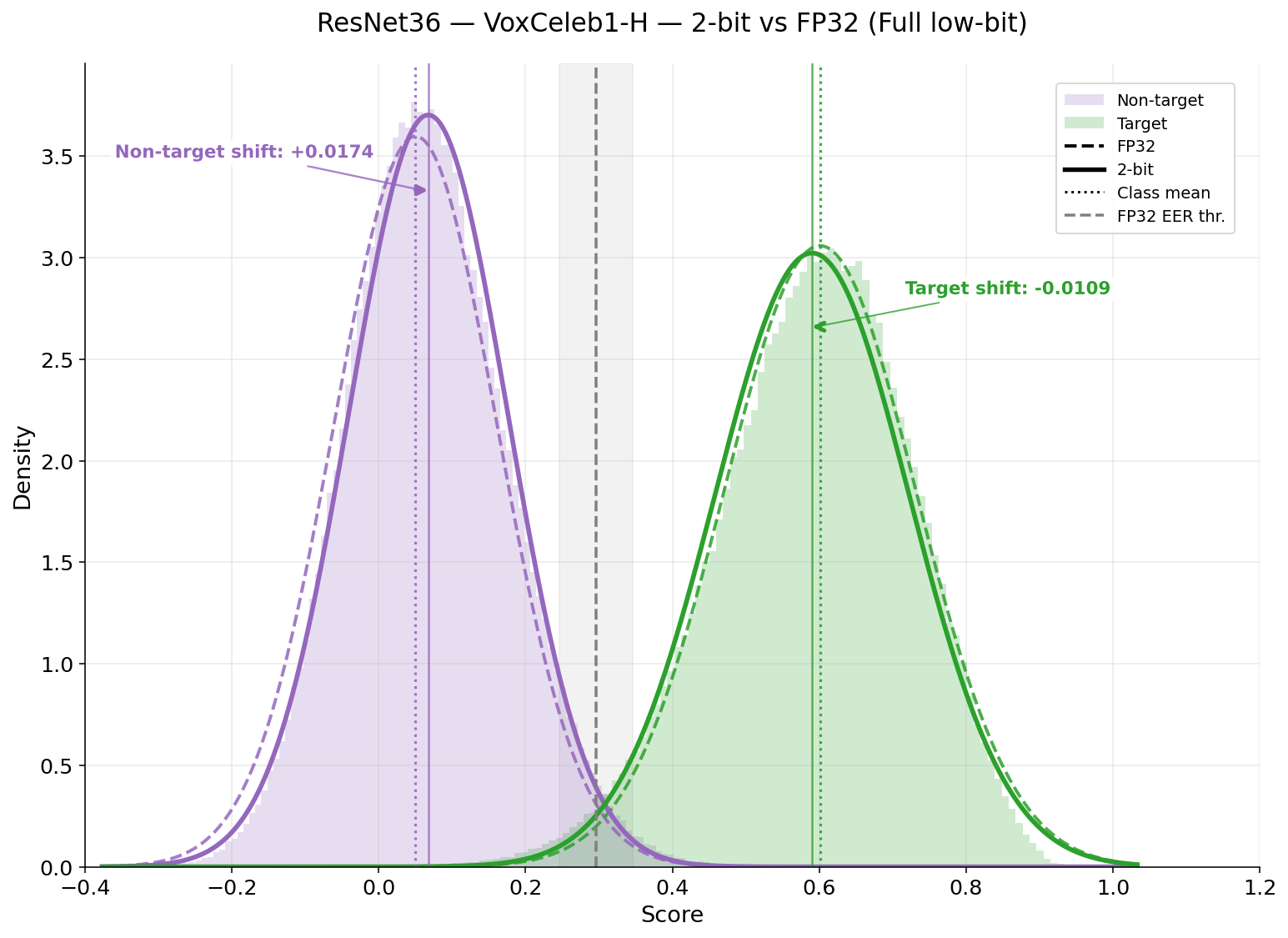}
    \caption{ResNet-36 score distributions on VoxCeleb1-H in the full low-bit setting. Dashed: FP32; solid: 2-bit. Non-target scores drift upward and target scores drift downward.}
    \label{fig:r36_2bit_fp32_shift_vox1h}
\end{figure}

Figure~\ref{fig:r36_2bit_fp32_shift_vox1h} illustrates this phenomenon on VoxCeleb1-H for ResNet-36 in the full low-bit setting. Compared with FP32, 2-bit quantization shifts non-target scores upward and target scores downward, which contracts the target/non-target gap and increases distribution overlap around the FP32 EER threshold. This visualization makes clear why low-bit quantization produces more fragile decisions, especially for trials lying close to the operating threshold.

A second observation is that the test-only drift remains relatively close to the full drift across all settings. This indicates that quantizing the test side alone already accounts for a large fraction of the total score perturbation. The effect is particularly strong for ResNet-36 at 2 bits, where the mean absolute test-only drift nearly matches the full drift ($0.0277$ vs. $0.0301$). Overall, these results show that low-bit quantization progressively destabilizes the score space as the bit-width decreases, with the strongest effect appearing in the 2-bit regime.

\subsection{Harmful flips concentrate near the FP32 threshold}
\label{sec:harmful}
While average score drift is informative, its practical impact depends on how close a trial lies to the operating threshold. To study this, we measure the distance of each FP32 score to the FP32 decision threshold. For a trial $t$, we define the FP32 margin as:
\begin{equation}
m(t) = |s_{ff}(t) - \tau^\star_{\mathcal S}|,
\label{eq:margin}
\end{equation}
where $s_{ff}(t)$ is the FP32 score and $\tau^\star_{\mathcal S}$ is the FP32 EER threshold estimated independently for subset $\mathcal S$. Intuitively, this margin reflects how close the trial is to the FP32 decision boundary under the FP32 system: trials with small $m(t)$ lie close to the decision boundary and are therefore ambiguous, whereas trials with large $m(t)$ are far from the threshold and are more decisively classified.

We then analyze the probability of a \emph{harmful flip}, i.e., the probability that quantization changes the binary decision obtained with the FP32 system in a harmful way, as a function of this margin. The key idea is that a given score drift matters only relative to the trial’s distance to the threshold: even a moderate perturbation can flip the decision of a near-threshold trial, while the same perturbation will usually have no effect on a trial that lies far from the boundary.

Table~\ref{tab:margin_harmful} shows that harmful flips are highly concentrated near the FP32 threshold. This effect is particularly clear on VoxCeleb1-H. For ResNet-36 at 2 bits, the harmful-flip rate is $0.2389$ for trials in the closest bin, $m \in [0,0.01)$, meaning that nearly one out of four highly ambiguous trials changes decision under quantization. However, this rate drops to $0.0335$ for $m \in [0.05,0.1)$ and to only $1.6\times10^{-4}$ for $m \ge 0.1$. ResNet-200 exhibits the same behavior, with a harmful-flip rate of $0.2204$ in the closest bin and $5.8\times10^{-5}$ beyond $0.1$. Similar patterns are observed at 3 and 4 bits, although at lower levels overall.

\begin{table}[H]
\centering
\tiny
\caption{Harmful-flip rate conditioned on the FP32 score margin $m=|s_{ff}-\tau^\star_{\mathcal S}|$, reported here on VoxCeleb1-H. Each cell gives the fraction of trials in the corresponding margin bin for which the quantized system changes the FP32 binary decision in a harmful direction with respect to the trial label.}
\label{tab:margin_harmful}
\resizebox{\columnwidth}{!}{
\begin{tabular}{lcccc}
\toprule
\textbf{Model} & \textbf{Bits} & \textbf{[0,0.01)} & \textbf{[0.05,0.1)} & \textbf{[0.1,+$\infty$)} \\
\midrule
ResNet-36  & 4 & 0.2043 & 0.0014 & 2.0 \texttimes{} 10\textsuperscript{-6} \\
ResNet-36  & 3 & 0.2138 & 0.0037 & 5.8 \texttimes{} 10\textsuperscript{-6} \\
ResNet-36  & 2 & 0.2389 & 0.0335 & 1.6 \texttimes{} 10\textsuperscript{-4} \\
\midrule
ResNet-200 & 4 & 0.2181 & 0.0093 & 9.6 \texttimes{} 10\textsuperscript{-6} \\
ResNet-200 & 3 & 0.2115 & 0.0137 & 2.3 \texttimes{} 10\textsuperscript{-5} \\
ResNet-200 & 2 & 0.2204 & 0.0241 & 5.8 \texttimes{} 10\textsuperscript{-5} \\
\bottomrule
\end{tabular}
}
\end{table}

These results show that quantization-induced decision errors are not uniformly distributed over the score space. Instead, they are strongly localized around the operating threshold, where the FP32 system is already least certain. Trials far from the threshold are rarely destabilized, even in the 2-bit regime, whereas near-threshold trials are much more vulnerable to low-bit score perturbations. This is an important practical result: it suggests that the majority of trials can be processed safely at low precision, while only a relatively narrow band of ambiguous cases requires additional care.

%%%%%%%%%%%%%%%%%%%%%%%%%%%%%%%%%%%%%%%%%%%%%%%%%%%%%%%%%%%%%%%%%%%%%%%%
%%%%%%%%%%%%%%%%%%%%%%%%%%% Section VI %%%%%%%%%%%%%%%%%%%%%%%%%%%%%%%%
%%%%%%%%%%%%%%%%%%%%%%%%%%%%%%%%%%%%%%%%%%%%%%%%%%%%%%%%%%%%%%%%%%%%%%%%

\section{Calibrated Multi-Precision Cascade}
\label{sec:Cascading}

In this section, we introduce a multi-precision cascade where trials are first handled at at 2 bits and only ambiguous cases are escalated to higher precision. First we present a calibrated gating rule based on the distance of low-bit scores to the FP32 threshold (Section~\ref{sec:gating}), then evaluate its verification performance and average sequential precision cost (Section~\ref{sec:cascade_perf}), and finally compare its memory footprint with single-model mixed-precision deployment based on MSFT~\cite{Liu_2024_KMQAT} (Section~\ref{sec:memory_msft}).

\begin{table*}[t]
\centering
\caption{Speaker verification performance of FP32, uniformly 2-bit quantized, MSFT, and cascade models on in-domain and out-of-domain benchmarks. Bold values indicate the best compressed system for each metric, excluding the FP32 reference.}
\label{tab:quant_results_cascade}
\renewcommand{\arraystretch}{1.0}
\setlength{\tabcolsep}{4pt}

\begin{adjustbox}{width=\textwidth}
\begin{tabular}{@{}cccc|cccccccc|cccccc@{}}
\toprule
\multirow{3}{*}{\textbf{System}} & \multirow{3}{*}{\textbf{\begin{tabular}[c]{@{}c@{}}Bit Width\\ (bit)\end{tabular}}} & \multicolumn{2}{c}{\textbf{Overall}} & \multicolumn{8}{c}{\textbf{In-Domain}} & \multicolumn{6}{c}{\textbf{Out-of-Domain}} \\ \cmidrule(l){3-18} 
 &  & \multicolumn{2}{c}{\textbf{Avg.}} & \multicolumn{2}{c}{\textbf{Avg.}} & \multicolumn{2}{c}{\textbf{Vox1-O}} & \multicolumn{2}{c}{\textbf{Vox1-E}} & \multicolumn{2}{c}{\textbf{Vox1-H}} & \multicolumn{2}{c}{\textbf{Avg.}} & \multicolumn{2}{c}{\textbf{CommonBench}} & \multicolumn{2}{c}{\textbf{CN-Celeb}} \\ \cmidrule(l){3-18} 
 &  & \textbf{EER} & \textbf{minDCF} & \textbf{EER} & \textbf{minDCF} & \textbf{EER} & \textbf{minDCF} & \textbf{EER} & \textbf{minDCF} & \textbf{EER} & \textbf{minDCF} & \textbf{EER} & \textbf{minDCF} & \textbf{EER} & \textbf{minDCF} & \textbf{EER} & \textbf{minDCF} \\ \midrule
\multirow{6}{*}{ResNet-36} & 32 & 3.910 & 0.254 & 1.234 & 0.118 & 0.973 & 0.085 & 0.994 & 0.102 & 1.736 & 0.167 & 7.923 & 0.458 & 3.922 & 0.360 & 11.923 & 0.555 \\
 & 4 & \multicolumn{1}{l}{3.949} & \textbf{0.263} & \multicolumn{1}{l}{1.278} & 0.124 & \multicolumn{1}{l}{0.994} & 0.0856 & \multicolumn{1}{l}{1.042} & \textbf{0.110} & \multicolumn{1}{l}{1.799} & 0.175 & \multicolumn{1}{l}{7.955} & 0.471 & \multicolumn{1}{l}{4.077} & 0.376 & \textbf{11.833} & \textbf{0.566} \\
 & 3 & \multicolumn{1}{l}{3.965} & \textbf{0.263} & \multicolumn{1}{l}{1.297} & \textbf{0.123} & \multicolumn{1}{l}{0.999} & \textbf{0.0794} & \multicolumn{1}{l}{1.053} & 0.113 & \multicolumn{1}{l}{1.840} & 0.176 & \multicolumn{1}{l}{7.966} & 0.473 & \multicolumn{1}{l}{4.078} & 0.379 & 11.855 & 0.568 \\
 & 2 & 4.292 & 0.287 & 1.462 & 0.146 & 1.068 & 0.1102 & 1.229 & 0.135 & 2.089 & 0.193 & 8.538 & 0.499 & 4.568 & 0.407 & 12.509 & 0.592 \\
 & MSFT & 3.969 & 0.265 & 1.278 & 0.128 & \textbf{0.986} & 0.091 & \textbf{1.036} & 0.116 & 1.811 & 0.177 & 8.006 & 0.471 & 4.090 & 0.376 & 11.922 & 0.567 \\
 & Cascade & \textbf{3.947} & \textbf{0.263} & \textbf{1.277} & 0.125 & 0.994 & 0.089 & 1.042 & 0.112 & \textbf{1.795} & \textbf{0.173} & \textbf{7.952} & \textbf{0.470} & \textbf{4.066} & \textbf{0.374} & 11.838 & 0.567 \\ \bottomrule
\end{tabular}
\end{adjustbox}
\end{table*}

\subsection{Calibrated multi-precision gating}
\label{sec:gating}

Let $s_b(t)$ denote the score of trial $t$ obtained with the $b$-bit model, and let $s_{\mathrm{fp32}}(t)$ denote the corresponding FP32 score. To compare distances to the decision boundary across precisions, we calibrate each low-bit score using a monotone isotonic mapping:
\begin{equation}
\hat{s}_b(t) = g_b\bigl(s_b(t)\bigr),
\label{eq:isotonic_calib}
\end{equation}
where $g_b$ is learned on a development set and then kept fixed for all evaluation subsets so that $\hat{s}_b$ matches the FP32 score scale. Let $\tau_{\mathrm{fp32}}$ denote the FP32 operating threshold estimated on the same development set. This calibration affects routing, but its monotonicity preserves score ordering; under domain shift, mismatch mainly increases escalation to higher precision. We then define the calibrated distance to threshold as
\begin{equation}
d_b(t) = \left| \hat{s}_b(t) - \tau_{\mathrm{fp32}} \right|.
\label{eq:dist_to_tau}
\end{equation}

Given two gating thresholds $\delta_2$ and $\delta_3$, selected on VoxTube~\cite{yakovlev23_interspeech} to optimize the EER--cost trade-off, the cascade operates as follows:
\begin{itemize}
    \item a trial is resolved at 2 bits if $d_2(t) > \delta_2$;
    \item it is re-scored at 3 bits and resolved there if $d_3(t) > \delta_3$;
    \item otherwise, it is re-scored at 4 bits.
\end{itemize}

This rule directly exploits the score structure identified in Section~\ref{sec:score_analysis}: when a calibrated low-bit score is sufficiently far from the FP32 threshold, the decision is typically robust, whereas only near-threshold trials require higher precision. The practical feasibility of this strategy is further supported by the strong inter-precision score correlations observed on ResNet-36, with Pearson correlation between 2-bit and FP32 scores reaching $0.992$ on VoxCeleb1-H and $0.986$ on VoxTube. 

\subsection{Cascade performance and average precision cost}
\label{sec:cascade_perf}

We use VoxTube~\cite{yakovlev23_interspeech} as the calibration set in order to avoid calibrating and evaluating on related VoxCeleb subsets. Table~\ref{tab:cascade_results} reports the routing statistics of the learned cascade, while Table~\ref{tab:quant_results_cascade} compares its verification performance with FP32, uniform 2-bit quantization, and MSFT.

The main result is that the cascade remains close to FP32 performance while improving over uniform 2-bit quantization. On the global average, it achieves 3.947\% EER, compared with 3.910\% for FP32 and 4.293\% for the pure 2-bit model. The same trend is observed both in-domain (1.277\% vs. 1.234\% for FP32 and 1.462\% for 2 bits) and out-of-domain (7.952\% vs. 7.923\% and 8.538\%, respectively). Overall, the cascade recovers most of the performance gap between uniform 2-bit quantization and FP32.

To summarize the effective computation of the cascade, we use a single sequential cost that reflects the actual inference path. In our setting, every trial is first scored at 2 bits, ambiguous trials are then re-scored at 3 bits, and only the most uncertain are finally re-scored at 4 bits. If $p_2$, $p_3$, and $p_4$ denote the fractions of trials ultimately resolved at 2, 3, and 4 bits, respectively, with $p_2+p_3+p_4=1$, then the average sequential cost is
\begin{equation}
c_{\mathrm{seq}} = 2 + 3(p_3+p_4) + 4p_4.
\label{eq:bseq}
\end{equation}
This expression directly reflects the cascade: all trials pay one 2-bit pass, the fraction $(p_3+p_4)$ additionally pays one 3-bit pass, and the fraction $p_4$ further pays one 4-bit pass.

With VoxTube calibration on the ResNet-36, this yields an average sequential cost of approximately $c_{\mathrm{seq}} \approx 2.66$ bits/trial on VoxCeleb1-E and $\approx3.11$ bits/trial on VoxCeleb1-H.

\begin{table}[htbp]
\centering
\small
\caption{Cascade routing statistics learned on VoxTube. Resolution rates are reported in percent. The Average Bit column gives the average sequential precision cost per trial, computed as $2 + 3((\%3b+\%4b)/100) + 4(\%4b/100)$.}
\label{tab:cascade_results}
\resizebox{\columnwidth}{!}{
\begin{tabular}{llcccc}
\toprule
\multirow{2}{*}{\textbf{Model}} & \multirow{2}{*}{\textbf{Subset}} & \multicolumn{3}{c}{\textbf{Resolution rate (\%)}} & \multirow{2}{*}{\textbf{Average Bit}} \\
\cmidrule(lr){3-5}
& & \textbf{2b} & \textbf{3b} & \textbf{4b} & \\
\midrule
ResNet-36  & VoxCeleb1-O & 86.05 & 7.85  & 6.09  & 2.66 \\
ResNet-36  & VoxCeleb1-E & 85.39 & 8.89  & 5.71  & 2.67 \\
ResNet-36  & VoxCeleb1-H & 77.63 & 11.14 & 11.21 & 3.12 \\
ResNet-36  & CN-Celeb    & 56.96 & 17.71 & 25.32 & 4.30 \\
ResNet-36  & CommonBench & 71.25 & 13.97 & 14.77 & 3.45 \\
%\midrule
%ResNet-200 & VoxCeleb1-E & 90.49 & 3.86  & 5.64  & 2.51 \\
%ResNet-200 & VoxCeleb1-H & 83.44 & 5.41  & 11.14 & 2.94 \\
%ResNet-200 & CN-Celeb    & 64.86 & 11.30 & 23.82 & 4.01 \\
%ResNet-200 & CommonBench & 77.80 & 7.38  & 14.80 & 3.26 \\
\bottomrule
\end{tabular}
}
\end{table}

\subsection{Memory footprint and comparison with MSFT}
\label{sec:memory_msft}

While the cascade reduces the average sequential precision budget, it does not necessarily minimize resident model memory. To clarify this trade-off, we distinguish between checkpoint storage and a deployable packed representation. In the latter, quantization indices are packed in $b$ bits, codebooks are stored in FP16, biases in FP16, and non-quantized parameters (e.g., batch-normalization parameters and running statistics) remain in FP32.

Under these assumptions, the deployable memory of a quantized model can be written as:

\begin{equation}
\mathrm{Mem}_{\mathrm{deploy}}
=
\sum_{\ell} N_\ell b_\ell
+
q_{\mathcal C}\sum_{\ell} |\mathcal{C}_\ell|
+
\mathrm{Mem}_{\mathrm{other}},
\label{eq:mem_deploy_simple}
\end{equation}

\noindent where $N_\ell$ denotes the number of quantized weights in layer $\ell$, $b_\ell$ its bit-width, $|\mathcal{C}_\ell|$ the codebook size, and $\mathrm{Mem}_{\mathrm{other}}$ groups bias terms and non-quantized parameters. We use $q_{\mathcal C}=16$ bits under the FP16 codebook assumption. In our implementation, the effective codebook already includes the learned scale, so no separate storage term is required for $\alpha$.

Table~\ref{tab:memory_msft} reports the resulting deployable memory for ResNet-36 and compression ratio CR$_w$. Uniform 2-bit quantization yields the smallest packed model, followed by uniform 3-bit, then MSFT, and finally uniform 4-bit. In this view, MSFT behaves as a compact single-model alternative close to uniform 3-bit quantization, with an average precision of $\bar b \simeq 3.144$ bits/weight.

This highlights a key trade-off. A cascade implemented with three resident models (2b, 3b, and 4b) is attractive for reducing average computation per trial, but not necessarily for minimizing memory footprint, since the combined storage is much larger than a single MSFT model. Conversely, MSFT is more favorable when deployment is primarily constrained by model storage. The two approaches therefore address different resource constraints: cascade for average inference cost, and MSFT for compact single-model deployment.

\begin{table}[t]
\centering
\small
\caption{Deployable memory footprint under packed-index assumptions.}
\label{tab:memory_msft}
\resizebox{\columnwidth}{!}{
\begin{tabular}{lcccccc}
\toprule
\textbf{Model} & \textbf{Bit Width} & \textbf{Avg. bits} & \textbf{Deploy (MB)} & \textbf{Packed (MB)} & \textbf{Codebooks (MB)} & \textbf{CR$_w$} \\
\midrule
ResNet-36 & FP32         & 32.0 & 62.89 & --   & --    & -- \\
ResNet-36 & 4   & 4.0  & 13.54 & 7.10 & 0.371 & $\times 4.64$ \\
ResNet-36 & 3   & 3.0  & 11.57 & 5.33 & 0.186 & $\times 5.43$ \\
ResNet-36 & 2   & 2.0  & 9.71  & 3.55 & 0.093 & $\times 6.47$ \\
ResNet-36 & MSFT (2/3/4) & 3.144  & 11.86 & 5.58 & 0.214 & $\times 5.30$ \\
ResNet-36 & Cascade      & 3.240  & 34.82 & 15.98 & 0.650 & $\times 1.80$ \\
\bottomrule
\end{tabular}
}
\end{table}

Overall, the cascade supports a simple principle: because low-bit score errors are strongly localized near the FP32 threshold and scores remain well correlated across precisions, most trials can be handled at 2 bits and only a small ambiguous subset requires higher precision. This substantially reduces the average computation of inference, while leaving open a complementary trade-off with single-model mixed-precision deployment.

\section{Conclusion}
\label{sec:Conclusion}
This paper examines low-bit KMQAT for speaker verification on ResNet-36 and ResNet-200 with uniform 4-bit, 3-bit, and 2-bit quantization, across in-domain and out-of-domain benchmarks. The results show that 2-bit quantization causes the largest performance drop, while 3- and 4-bit models remain close to FP32. 
Stage-wise retention experiments reveal that degradation is not uniform across the network and is mainly associated with middle-to-late layers. At the score level, low-bit errors are structured: as precision decreases, score drift becomes much stronger, notably at 2 bits, and harmful flips concentrate near the FP32 threshold. 
To address this, a calibrated multi-precision cascade was proposed. It uses low-bit inference for easy cases and higher precision only for ambiguous trials. 
Overall, the study shows that low-bit quantization in speaker verification should be evaluated not only with global accuracy metrics, but also by analyzing score perturbations, decision flips, and the trade-off between inference cost and model memory.

\section{Acknowledgements}

This project was provided with computing HPC and storage resources by GENCI at IDRIS thanks to the grant 2026-AD011016050R1 on the supercomputer Jean Zay V100 partition.

\bibliographystyle{IEEEtran}
\bibliography{Odyssey2026_BibEntries}

@inproceedings{desplanques20_interspeech,
  title     = {{ECAPA-TDNN: Emphasized Channel Attention, Propagation and Aggregation in TDNN Based Speaker Verification}},
  author    = {Brecht Desplanques and Jenthe Thienpondt and Kris Demuynck},
  year      = {2020},
  booktitle = {{Interspeech}},
  pages     = {3830--3834},
  doi       = {10.21437/Interspeech.2020-2650},
  issn      = {2958-1796},
}

@article{zeinali2019but,
  title={But system description to voxceleb speaker recognition challenge 2019},
  author={Zeinali, Hossein and Wang, Shuai and Silnova, Anna and Mat{\v{e}}jka, Pavel and Plchot, Old{\v{r}}ich},
  journal={arXiv preprint arXiv:1910.12592},
  year={2019}
}

@inproceedings{yakovlev24_interspeech,
  title     = {Reshape Dimensions Network for Speaker Recognition},
  author    = {Ivan Yakovlev and Rostislav Makarov and Andrei Balykin and Pavel Malov and Anton Okhotnikov and Nikita Torgashov},
  year      = {2024},
  booktitle = {Interspeech 2024},
  pages     = {3235--3239},
  doi       = {10.21437/Interspeech.2024-2116},
}

@inproceedings{thienpondt2023ecapa2,
  title={{ECAPA2}: A hybrid neural network architecture and training strategy for robust speaker embeddings},
  author={Thienpondt, Jenthe and Demuynck, Kris},
  booktitle={2023 IEEE automatic speech recognition and understanding workshop (ASRU)},
  pages={1--8},
  year={2023},
  organization={IEEE}
}

@inproceedings{snyder2018x,
  title={{X-Vectors}: Robust dnn embeddings for speaker recognition},
  author={Snyder, David and Garcia-Romero, Daniel and Sell, Gregory and Povey, Daniel and Khudanpur, Sanjeev},
  booktitle={2018 IEEE international conference on acoustics, speech and signal processing (ICASSP)},
  pages={5329--5333},
  year={2018},
  organization={IEEE}
}

@misc{lin2022lightweightapplicationsasymmetricenrollverify,
      title={Towards Lightweight Applications: Asymmetric Enroll-Verify Structure for Speaker Verification}, 
      author={Qingjian Lin and Lin Yang and Xuyang Wang and Xiaoyi Qin and Junjie Wang and Ming Li},
      year={2022},
      eprint={2110.04438},
      archivePrefix={arXiv},
      primaryClass={cs.SD},
      url={https://arxiv.org/abs/2110.04438}, 
}

@INPROCEEDINGS{wang2019kd_smallfootprint,
  author={Wang, Shuai and Yang, Yexin and Wang, Tianzhe and Qian, Yanmin and Yu, Kai},
  booktitle={ICASSP 2019 - 2019 IEEE International Conference on Acoustics, Speech and Signal Processing (ICASSP)}, 
  title={Knowledge Distillation for Small Foot-print Deep Speaker Embedding}, 
  year={2019},
  volume={},
  number={},
  pages={6021-6025},
  doi={10.1109/ICASSP.2019.8683443}}

@inproceedings{li23t_interspeech,
  author    = {Jingyu Li, Wei Liu and Zhaoyang Zhang and Jiong Wang and Tan Lee},
  title     = {{Model Compression for DNN-based Speaker Verification Using Weight Quantization}},
  year      = {2023},
  booktitle = {{Interspeech 2023}},
  pages     = {1988--1992},
  doi       = {10.21437/Interspeech.2023-1524},
  issn      = {2958-1796}
}

@INPROCEEDINGS{liu2022self_kd,
  author={Liu, Bei and Wang, Haoyu and Chen, Zhengyang and Wang, Shuai and Qian, Yanmin},
  booktitle={ICASSP 2022 - 2022 IEEE International Conference on Acoustics, Speech and Signal Processing (ICASSP)}, 
  title={Self-Knowledge Distillation via Feature Enhancement for Speaker Verification}, 
  year={2022},
  volume={},
  number={},
  pages={7542-7546},
  doi={10.1109/ICASSP43922.2022.9746529}}

@inproceedings{wang23u_interspeech,
  title     = {{Adaptive Neural Network Quantization For Lightweight Speaker Verification}},
  author    = {Haoyu Wang and Bei Liu and Yifei Wu and Yanmin Qian},
  year      = {2023},
  booktitle = {{Interspeech 2023}},
  pages     = {5331--5335},
  doi       = {10.21437/Interspeech.2023-927},
  issn      = {2958-1796},
}

@article{Liu_2024_KMQAT,
   title={Towards Lightweight Speaker Verification via Adaptive Neural Network Quantization},
   volume={32},
   ISSN={2329-9304},
   url={http://dx.doi.org/10.1109/TASLP.2024.3437237},
   DOI={10.1109/taslp.2024.3437237},
   journal={IEEE/ACM Transactions on Audio, Speech, and Language Processing},
   publisher={Institute of Electrical and Electronics Engineers (IEEE)},
   author={Liu, Bei and Wang, Haoyu and Qian, Yanmin},
   year={2024},
   pages={3771–-3784} }

@misc{bengio2013STE,
      title={Estimating or Propagating Gradients Through Stochastic Neurons for Conditional Computation}, 
      author={Yoshua Bengio and Nicholas L{\'e}onard and Aaron Courville},
      year={2013},
      eprint={1308.3432},
      archivePrefix={arXiv},
      primaryClass={cs.LG},
      url={https://arxiv.org/abs/1308.3432}, 
}

@inproceedings{rouvier2026, 
   title={{Kiwano: A Cutting-Edge Open-Source Toolkit for Speaker Verification}},
   booktitle={Odyssey 2026},
   author={Rouvier, Mickael and Bousquet, Pierre-Michel},
   year={2026},
    }

@inproceedings{chung2018voxceleb2,
  author    = "J. S. Chung and A. Nagrani and A. Zisserman",
  title     = {{VoxCeleb2}: Deep Speaker Recognition},
  booktitle = "Interspeech",
  year      = "2018",
  note      = "arXiv:1806.05622"
}

@InProceedings{Nagrani17,
	author       = "Nagrani, A. and Chung, J.~S. and Zisserman, A.",
	title        = {{VoxCeleb}: a large-scale speaker identification dataset},
	booktitle    = "Interspeech",
	year         = "2017",
}

@inproceedings{hintz24_spsc,
  title     = {{CommonBench}: A larger Scale Speaker Verification Benchmark},
  author    = {Jan Hintz and Ingo Siegert},
  year      = {2024},
  booktitle = {4th Symposium on Security and Privacy in Speech Communication},
  pages     = {17--20},
  doi       = {10.21437/SPSC.2024-3},
}

@inproceedings{fan2019cnceleb,
  author    = "Y. Fan and L. Chen and S. Kang and et al.",
  title     = {{CN-Celeb}: A Challenging Chinese Speaker Recognition Dataset},
  booktitle = "Proc. Interspeech",
  year      = "2019"
}

@misc{snyder2015musanmusicspeechnoise,
      title={{MUSAN}: A Music, Speech, and Noise Corpus}, 
      author={David Snyder and Guoguo Chen and Daniel Povey},
      year={2015},
      eprint={1510.08484},
      archivePrefix={arXiv},
      primaryClass={cs.SD},
      url={https://arxiv.org/abs/1510.08484}, 
}

@INPROCEEDINGS{rirs,
  author={Ko, Tom and Peddinti, Vijayaditya and Povey, Daniel and Seltzer, Michael L. and Khudanpur, Sanjeev},
  booktitle={International Conference on Acoustics, Speech and Signal Processing (ICASSP)}, 
  title={A study on data augmentation of reverberant speech for robust speech recognition}, 
  year={2017},
  volume={},
  number={},
  pages={5220-5224},
  doi={10.1109/ICASSP.2017.7953152}}

@misc{rouvier2021studyingsqueezeandexcitationusedcnn,
      title={Studying squeeze-and-excitation used in CNN for speaker verification}, 
      author={Mickael Rouvier and Pierre-Michel Bousquet},
      year={2021},
      eprint={2109.05977},
      archivePrefix={arXiv},
      primaryClass={eess.AS},
      url={https://arxiv.org/abs/2109.05977}, 
}

@inproceedings{yakovlev23_interspeech,
  title     = {{VoxTube: a multilingual speaker recognition dataset}},
  author    = {Ivan Yakovlev and Anton Okhotnikov and Nikita Torgashov and Rostislav Makarov and Yuri Voevodin and Konstantin Simonchik},
  year      = {2023},
  booktitle = {{Interspeech 2023}},
  pages     = {2238--2242},
  doi       = {10.21437/Interspeech.2023-1083},
  issn      = {2958-1796}
}

\end{document}